\documentclass[prd,aps,eqsecnum,amsmath,floatfix,nofootinbib,preprint,tightenlines]{revtex4}

\usepackage{latexsym}
\usepackage{graphicx}
\usepackage{multirow}
\usepackage[dvipsnames]{xcolor}
\usepackage{hyperref}

\def\NON{\nonumber\\}
\def\bibi{\bibitem}

\def\a{\alpha}

\def\c{\chi}
\def\d{\delta}
\def\g{\gamma}

\def\j{\psi}

\def\l{\lambda}
\def\m{\mu}
\def\n{\nu}

\def\p{\pi}                     
\def\th{\theta}                  

\def\x{\xi}
\def\z{\zeta}
\def\D{\Delta}
\def\F{\Phi}
\def\G{\Gamma}
\def\J{\Psi}

\def\S{\Sigma}

\def\cl{{\cal L}}

\def\bo{\raisebox{-.4ex}{\large$\Box$}}                 
\def\cbo{{\,\raise-.15ex\Sc [\,}}                       

\def\Sl#1{\rlap{\hbox{$\mskip 3 mu /$}}#1}      
\def\bra#1{\Big\langle #1\Big|}                 
\def\ket#1{\Big| #1\Big\rangle}                 
\def\vev#1{\Big\langle #1 \Big\rangle}           

\def\svev#1{\left\langle #1\right\rangle}       

\def\ddt#1{{\buildrel {\hbox{\LARGE .\kern-2pt.}} \over {#1}}}

\def\ie{\mbox{\it i.e.}}
\def\eg{\mbox{\it e.g.}}

\def\tr{{\rm tr}\,}

\def\half{{1\over 2}}

\def\det{{\rm det}}

\def\ttl#1{{\it #1}}

\def\vev#1{\langle #1\rangle}
\def\bra#1{\langle #1|}
\def\ket#1{|#1\rangle}

\def\bx{\overline{\xi}}
\def\bj{\overline{\j}}
\def\bJ{\overline{\J}}

\def\hU{\hat{U}}

\def\Dslash{\Sl{D}}

\def\cl{{\cal L}}

\def\svev#1{\left\langle #1\right\rangle}       
\def\Sl#1{\rlap{\hbox{$\mskip 3 mu /$}}#1}      

\def\tr{\mbox{tr}}

\def\vev#1{\langle#1\rangle}

\def\irrep{{\it irrep}}
\def\irreps{{\it irreps}}

\def\id{{\bf 1}}

\begin{document}

\begin{center}
{\large\bf One-loop Chiral Perturbation Theory
  with two fermion representations}\\[8mm]
Thomas DeGrand,$^a$\ Maarten Golterman,$^b$\ Ethan T.~Neil$^{a,c}$\
and Yigal Shamir$^d$\\[8 mm]
{\small
$^a$Department of Physics,
University of Colorado, Boulder, CO 80309, USA\\
$^b$Department of Physics and Astronomy, San Francisco State University,\\
San Francisco, CA 94132, USA\\
$^c$RIKEN-BNL Research Center, Brookhaven National Laboratory,
Upton, NY 11973, USA\\
$^d$Raymond and Beverly Sackler School of Physics and Astronomy,\\
Tel~Aviv University, 69978, Tel~Aviv, Israel}\\[10mm]
\end{center}

\begin{quotation}
We develop Chiral Perturbation Theory for chirally broken
theories with fermions in two different representations
of the gauge group.  Any such theory has
a non-anomalous singlet $U(1)_A$ symmetry, yielding an additional
Nambu--Goldstone boson when spontaneously broken.
We calculate the next-to-leading order corrections for the pseudoscalar
masses and decay constants, which include the singlet Nambu--Goldstone boson,
as well as for the two condensates.  The results can be generalized
to more than two representations.
\end{quotation}

\newpage
\section{\label{intro} Introduction}
Within the Standard Model, all of the quarks transform in the
fundamental representation of the QCD group $SU(3)_c$.
However, exotic fermions in higher irreducible representations (\irreps)
of $SU(3)_c$ have long been considered an intriguing possibility for
physics beyond the Standard Model, with a potentially rich phenomenology
\cite{Ma:1975qy,Karl:1976jk,Wilczek:1976qi,Ng:1978qt,Georgi:1979ng,
  Dover:1979sn,Marciano:1980zf,Holdom:1982ex, Konishi:1982xh, Lust:1985aw,
  Clark:1986vk,Braaten:1986eg, Borisov:1986ev, White:1987pq, Fukazawa:1990fb,
  Chivukula:1990di, Simmons:1990px, Chivukula:1991zk,Kahara:2012yr}.
More generally, fermions in multiple representations of a
strongly-coupled gauge group can appear in other
extensions of the Standard Model, including
composite Higgs \cite{Contino:2010rs, Bellazzini:2014yua,Panico:2015jxa}
and composite dark matter \cite{Kribs:2016cew} models.
In particular, composite Higgs models of ``partial compositeness"
\cite{Kaplan:1991dc}, in which the elementary top quark mixes with a
composite top partner, tend to require the presence of fermions in two
different representations of the new strongly-coupled gauge group
\cite{Ferretti:2013kya, Barnard:2013zea, Ferretti:2014qta,
  Golterman:2015zwa, Ferretti:2016upr}.

In any of the extensions of the Standard Model noted above,
the presence of strong gauge
interactions impedes the use of perturbation theory for most quantities
of interest.  If the strong sector exhibits spontaneous chiral symmetry
breaking, then the dynamics of the resulting Nambu-Goldstone bosons
(NGBs) can be described by a low-energy effective theory known as
Chiral Perturbation Theory (ChPT) \cite{Weinberg:1978kz,Gasser:1983yg,
  Gasser:1984gg,Golterman:2009kw}.  ChPT is an invaluable tool for
understanding the associated phenomenology, and has been used with great
success in the context of QCD.  Looking beyond QCD, ChPT is also
well-understood for the case of an arbitrary number of fermions in a
single representation, including complex \cite{Gasser:1986vb}
as well as real or pseudoreal representations
\cite{Splittorff:2001fy,Bijnens:2009qm,Bijnens:2011fm,Bijnens:2011xt}.
However, ChPT has not been systematically explored in the case of a
strong sector containing two or more fermion representations.

With fermions in two different representations, the chiral symmetry
breaking pattern remains mostly unchanged: if each fermion species $r$ in
isolation has an associated global chiral symmetry $G_r$ which is
spontaneously broken to $H_r$, then when multiple species are present
the global symmetry contains the product group $G_1 \times G_2 \times
\cdots \times G_n$, and the residual unbroken symmetry group is $H_1 \times
H_2 \times \cdots \times H_n$.  However, this is not the whole story; with
two or more fermion representations, additional abelian axial
symmetries appear as linear combinations of the individually
anomalous flavor-singlet axial rotations of each fermion species.
These additional symmetries are then spontaneously broken,
giving rise to singlet NGBs \cite{Clark:1986vk,White:1993ms}.

Any additional singlet NGB which appears in a theory with multiple
fermion representations is a particularly interesting object.  It can
play the role of a composite axion \cite{Kim:1984pt,Randall:1992ut,
  Rubakov:1997vp,Kobakhidze:2016wmv,Barrie:2016ntq},
offering a potential solution to the strong CP problem.
In various extensions of the Standard Model,
the singlet may provide a candidate to explain the 750 GeV
diphoton excess observed by ATLAS and CMS \cite{ATLAS:750,CMS:2015dxe};
within composite Higgs models it appears quite naturally as a relatively
isolated light state with anomaly-induced couplings to pairs of
Standard-Model vector bosons \cite{Belyaev:2015hgo, Ferretti:2016upr}.

In this paper, we study ChPT through next-to-leading order (NLO)
for a theory with two fermion species charged under distinct
representations of a confining gauge group; generalization to more than
two species is straightforward.  All fermion masses for a particular
representation are taken to be degenerate for simplicity.  We derive
formulas for the pseudoscalar masses and decay constants of all states,
including the singlet NGB.  We also give formulas for the two
chiral condensates.

The outline of the paper is as follows:
In Sec.~\ref{syms} we describe the symmetries and patterns of breaking for the
three types of \irreps.  While being well established, we found it useful
to include this discussion, to make this paper more self-contained.
In Sec.~\ref{ChL} we write down the chiral Lagrangian through order $p^4$
for a theory with two representations of fermions.
One-loop results for the pseudoscalar masses, decay constants,
and condensates are presented in Sec.~\ref{1loop}.  We conclude in Sec.~\ref{conc}.
The three appendices deal with technicalities.

\section{\label{syms} Symmetries and patterns of breaking}
There are  three types of \irreps: complex, real, and pseudoreal.
We consider a vector-like field content, which implies that fermions
in a complex or pseudoreal \irrep\ fit into $N$ Dirac fermions.
For a real \irrep, we will allow any number $N_w$ of Weyl (or, equivalently,
Majorana) fermions.  In the chiral limit, where all masses are zero,
the familiar symmetry breaking patterns are \cite{Peskin:1980gc}
\begin{eqnarray}
  {\rm complex}: & SU(N)_L \times SU(N)_R \rightarrow SU(N)_V \ ,
\label{symptrn}\\
  {\rm pseudoreal}: & SU(2N) \rightarrow Sp(2N) \ ,
\NON
  {\rm real}: & SU(N_w) \rightarrow SO(N_w) \ .
\nonumber
\end{eqnarray}
As a natural generalization of the familiar terminology of QCD,
for all types of \irreps\ we will refer to the spontaneously broken symmetries
as axial symmetries, and to the unbroken ones as vector symmetries.
For simplicity, we will consider only mass matrices that do not break
explicitly any of the vector symmetries, so that all pions made out of
a single fermion species will have the same mass.

\subsection{\label{pattern} Symmetry breaking patterns}
In this subsection we describe in some detail the symmetry breaking
pattern for each type of \irrep, and how it is reflected in the
field content of the effective chiral theory.

\subsubsection{\label{cmplxrep} Complex representations}
We consider $N$ Dirac fermions $\j_i,\bj_i,$ where the flavor index
is $i=1,\ldots,N$.  We suppress color and Dirac indices.
The global symmetry of the massless theory is $SU(N)_L\times SU(N)_R$,
which is spontaneously broken to the diagonal subgroup $SU(N)_V$.
The effective field $\S$ takes values in the
coset $SU(N)_L\times SU(N)_R/SU(N)_V\cong SU(N)$.
It describes the long-distance fluctuations of the bilinears
\begin{eqnarray}
  \S_{ij} & \leftrightarrow &
  \tr (P_L \j_i \bj_j) = \tr (\j_{L,i} \bj_{R,j}) \ ,
\label{cmplxSig}\\
  \S^*_{ij} & \leftrightarrow &
  \tr (P_R \j_j \bj_i) = \tr (\j_{R,j} \bj_{L,i}) \ ,
\nonumber
\end{eqnarray}
where the traces on the right-hand side
are over color and Dirac indices, $P_{R,L}=(1\pm\g_5)/2$,
and $\j_{L,R} = P_{L,R}\j$, $\bj_{L,R} = \bj P_{R,L}$.
The chiral spurion $\c_{ij}(x)$ is introduced by adding to the Lagrangian
of the massless theory the following source term
\begin{equation}
  \cl_{src} = \bj_L \c \j_R + \bj_R \c^\dagger \j_L \ .
\label{chicmplx}
\end{equation}
The symmetry transformations act as
\begin{subequations}
\label{cmplx}
\begin{eqnarray}
  \j_{L,R} &\to& g_{L,R}\, \j_{L,R} \ , \qquad
  \bj_{L,R} \ \to \ \bj_{L,R}\, g_{L,R}^\dagger \ ,
\label{transxa}\\
  \S &\to& g_L \,\S\, g_R^\dagger \ , \qquad
  \c \ \to \ g_L \,\c\, g_R^\dagger \ ,
\label{transxb}
\end{eqnarray}
\end{subequations}
where $g_{L,R} \in SU(N)_{L,R}$.

The mass matrix is given by the ``expectation value'' of the chiral spurion,
$M_{ij}=\svev{\c_{ij}}$.  By applying an $SU(N)_L\times SU(N)_R$ transformation
the mass term can be brought to a diagonal form, $M_{ij}=m_i \d_{ij}$,
where in general $m_i$ are complex numbers.
In this paper, we will consider only the equal-mass limit, $m_i=m$,
and we take $m$ to be real and positive.
The fermion condensate will therefore be oriented in the direction
of the identity matrix, $\svev{\bj_i\j_j} \propto \d_{ij}$.
Correspondingly, for the effective field we will have $\svev{\S_{ij}}=\d_{ij}$.

\subsubsection{\label{rprep} Real and pseudoreal representations}
For any real or pseudoreal \irrep\ there exists a matrix $S$ with
the invariance property
\begin{equation}
  g^T S g = S \ ,
\label{invS}
\end{equation}
for any element $g$ of the gauge group.
Here $S$ is a real orthogonal matrix.
Equivalently, the hermitian generators
of the Lie algebra $T_a$ satisfy
\begin{equation}
  T^T_a S = T^*_a S = -S T_a \ .
\label{genS}
\end{equation}
For a real representation $S$ is symmetric, whereas for a pseudoreal
representation it is antisymmetric.

We start by considering again $N$ Dirac fermions, and begin by studying
their properties under charge conjugation. The massless action
for any number of Dirac fermions in a complex \irrep\ is invariant
under charge conjugation, which acts on the fermion and gauge fields as
\begin{eqnarray}
  \j &\to & C\, \bj^T \ ,
\label{cconj}\\
  \bj  &\to& \j^T C \ ,
\NON
  A_\m &\to & -A_\m^*  \ ,
\nonumber
\end{eqnarray}
where the charge-conjugation matrix $C$ satisfies $C\g_\m = -\g_\m^T C$,
and $C^{-1}=C^\dagger = C^T = -C$.

For Dirac fermions that belong to a real or a pseudoreal \irrep,
the massless fermion action is invariant
under an additional, similar-looking discrete symmetry that
leaves the gauge field invariant, and acts non-trivially
on the fermion fields only, according to\footnote{%
  This is referred to as ``anti-unitary'' symmetry in Ref.~\cite{Kogut:2000ek}
}
\begin{eqnarray}
  \j &\to & S C\, \bj^T \ ,
\label{cconjrp}\\
  \bj  &\to& \j^T C S^T \ .
\nonumber
\end{eqnarray}
Because the gauge field is invariant, the transformation~(\ref{cconjrp})
can be applied to each Dirac fermion individually.

Motivated by this symmetry, we express the microscopic theory in terms
of purely left-handed Weyl fermions, $\x_I \equiv P_L\x_I\equiv\x_{L,I}$,
$\bx_I \equiv \bx_I P_R\equiv\bx_{L,I}$, where $I=1,\ldots,2N$,
which are related to the left- and right-handed components of
the Dirac fermions via\footnote{%
  Technically, we define the Weyl fermions as 4-component fields
  whose right-handed components vanish identically.
}
\begin{eqnarray}
\label{weylbasis}
  \x_{L,i} &=& \j_{L,i} \ ,
\\
  \x_{L,N+i} &=& SC\,\bj^T_{R,i} \ ,
\NON
  \bx_{L,i} &=& \bj_{L,i}\ ,\nonumber\\
  \bx_{L,N+i} &=& \j^T_{R,i}S^TC \ .
\nonumber
\end{eqnarray}
In terms of the Weyl fields, the Lagrangian takes the form
\begin{equation}
  \cl = \sum_{I=1}^{2N} \bx_{L,I} \Dslash\, \x_{L,I}\ ,
\label{lagweyl}
\end{equation}
which is invariant under the $SU(2N)$ flavor transformation
\begin{equation}
\x_L\to g\x_L\ ,\qquad \bx_L\to\bx_L g^\dagger\ .
\label{SU2Nw}
\end{equation}
Because of the Grassmann nature of the field, we have
\begin{equation}
  \x_{L,I}^T C S\x_{L,J} = \x_{L,J}^T C S^T \x_{L,I} \ ,
  \qquad \bx_{L,I} C S\bx^T_{L,J} = \bx_{L,J} C S^T \bx^T_{L,I}\ .
\label{bilchi}
\end{equation}
It follows that these bilinears are (anti)symmetric in $I\leftrightarrow J$
when $S$ is (anti)symmetric.

The chiral field $\S$ now lives in $SU(2N)$, with the correspondence
\begin{subequations}
\label{rpSig}
\begin{eqnarray}
  \tr (\x_{L,I} \x_{L,J}^T CS) & \leftrightarrow & \S_{IJ} = s\S_{JI} \ ,
\label{rpSiga}\\
  \tr (\bx_{L,I}^T \bx_{L,J} CS^T ) & \leftrightarrow & \S^*_{IJ} = s\S^*_{JI} \ ,
\label{rpSigb}
\end{eqnarray}
\end{subequations}
where it follows from Eq.~(\ref{bilchi}) that $s=1$ ($s=-1$) for a real (pseudoreal)
\irrep.  In both cases we have the transformation rules
\begin{equation}
  \S \to g \S g^T \ , \qquad \c \to g \c g^T \ ,
\label{transrpSig}
\end{equation}
and the source term in the Lagrangian is now
\begin{equation}
  \cl_{src} = \bx_L \c CS^T \bx_L^T + \x_L^T CS \c^\dagger \x_L \ .
\label{chirp}
\end{equation}

For a real \irrep, we will allow the number of Weyl fields $N_w$ to be
either even or odd.  In the latter case, one can then use a Weyl basis
or a Majorana basis (see App.~\ref{Majorana}), but not a Dirac basis.
For all values of $N_w$, we have that $\S$ is an element of the
coset generated by the broken generators, and thus an element of
$SU(N_w)$.

As usual, the symmetry-breaking order parameter is a fermion bilinear,
now given by the expectation value of Eq.~(\ref{bilchi}).  We will assume
that the mass matrix orients the fermion condensate such that
\begin{equation}
  \svev{\x_{L,J}^T \,C S\, \x_{L,I}} \propto J_{IJ} \ ,
\label{vevrp}
\end{equation}
where $J$ is a real orthogonal matrix, and where
$J$ is symmetric (antisymmetric) for a real (pseudoreal) \irrep.
While we will be making specific choices for the explicit form of
the matrix $J$, our discussion of the chiral effective theory
applies assuming only that $\svev{\S}=J$, where $J$ has
the properties listed above, and, in addition, $\det\, J = 1$.

For a pseudoreal \irrep, we will again assume that the Dirac mass matrix
is given by $M_{ij}=m\d_{ij}$, with $m\ge0$.  When translated to
the Weyl basis, the mass term takes the form
\begin{equation}
m\bj \j\to\half m\left(\x^T_LCSJ_A\x_L+\bx_L CSJ_A\bx^T_L\right)\ ,
\label{masstransl}
\end{equation}
with
\begin{equation}
  J_A = \left( \begin{array}{cc} 0 & -1 \\ 1 & 0 \end{array} \right)\ .
\label{JA}
\end{equation}
The fermion condensate is
oriented in the direction of the $2N\times 2N$ matrix $J_A$, and the
symmetry breaking pattern is $SU(2N)\to Sp(2N)$.
Note that $\det\, J_A = +1$, independent of $N$.  It follows
that the ground state is represented in the effective theory as
$\svev{\S_{IJ}}=(J_A)_{IJ}$, consistent with the fact that $\S\in SU(2N)$.

In the case of a real \irrep\ one can conceive of two simple choices
for the mass matrix.  First, for any number $N_w$
of Weyl (or Majorana) fermions, we may consider the
Majorana mass $M_{IJ}=m\d_{IJ}$, where again $m\ge0$.
The fermion condensate is then $\propto \d_{IJ}$, and the ground state
of the effective theory is $\svev{\S_{IJ}}=\d_{IJ}$.
By taking the chiral limit $m\to 0$, we see that the
symmetry breaking pattern is indeed $SU(N_w)\to SO(N_w)$.

In the case of an even number of Majorana fermions,
we may re-group the fields into $N=N_w/2$ Dirac fermions.
Let us endow these Dirac fermions with a common mass, $M_{ij}=m\d_{ij}$.
Upon translating back to the Weyl or Majorana basis,
the mass matrix takes the same form as in Eq.~(\ref{masstransl}),
except the $2N\times 2N$ matrix $J_A$ gets replaced by $J_S$, with
\begin{equation}
  J_S
  = \left( \begin{array}{cc} 0 & 1 \\ 1 & 0 \end{array} \right)\ .
\label{JS}
\end{equation}
Note, however, that
\begin{equation}
  \det\, J_S = \left\{
  \begin{array}{ll} +1\ , \qquad & \mbox{$N$ even} \ , \\
                    -1\ , \qquad & \mbox{$N$ odd} \ . \end{array}  \right.
\label{detJs}
\end{equation}
Only in the case that the number of Dirac fermions is even
(equivalently, the number of Majorana fermions is a multiple of 4)
is $J_S$ an element of $SU(2N)$, so that we may assume that
$\svev{\S_{IJ}}=(J_S)_{IJ}$.  In the case of an odd number of Dirac fermions,
it is in general not possible to have $\svev{\S}=J_S$.
This elementary fact is sometimes overlooked in the literature.

\subsection{\label{coset} Parametrization of the coset field: single \irrep}
In the case of a complex \irrep\ we have $\svev{\S_{ij}}=\d_{ij}$.
The expansion around this classical vacuum is facilitated by writing
$\S(x)=\hU(x)\in SU(N)$, with
\begin{equation}
  \hU(x) = \exp\left(\frac{i\sqrt{2}\pi(x)}{F}\right)
  = \exp\left(\frac{i\sqrt{2}\pi_a(x) T_a}{F}\right) \ ,
\label{Uhatc}
\end{equation}
with $\p_a$ the Nambu--Goldstone bosons (NGBs) associated with the
spontaneous symmetry breaking, and
where $T_a$, $a=1,\ldots,N^2-1$, are the generators of $SU(N)$,
normalized as\footnote{%
  The same normalization is used for the real and pseudoreal cases.
}
\begin{equation}
  \tr (T_a T_b) = \d_{ab} \ ,
\label{tnorm}
\end{equation}
and $F$ is the pion decay constant in the chiral limit.
Following Ref.~\cite{Bijnens:2009qm} we adopt the convention
\begin{equation}
\bra{0} A_{\m a}(x) \ket{\pi_b}
  = ip_\m \sqrt{2} F\, \d_{ab}\,e^{ipx}\, \ ,
\label{fpidef}
\end{equation}
where $A_\m^a$ is the axial current.
Introducing the (external) vector gauge field $v_\m=v_{\m a}(x) T_a$ and
the axial gauge field $a_\m(x) = a_{\m a}(x) T_a$, the covariant derivative
takes the form
\begin{equation}
  D_\m \S = D_\m \hU = \partial_\m \hU + i[v_\m,\hU] + i\{a_\m,\hU\} \ .
\label{corderc}
\end{equation}

Moving on to real and pseudoreal \irreps, we first split the generators
of the global symmetry group $SU(N_w)$
into broken generators $X_{\hat{a}}$ and unbroken generators
$Q_{\tilde{a}}$, which satisfy
\begin{eqnarray}
   J Q_{\tilde{a}}&=& - Q_{\tilde{a}}^T J \ ,
\label{unbrkrp}\\
   J X_{\hat{a}}&=& + X_{\hat{a}}^T J  \ ,
\label{brkrp}
\end{eqnarray}
where the matrix $J$ was introduced in Eq.~(\ref{vevrp}).  In both cases,
the expansion of the non-linear field can be written as
\begin{equation}
  \S(x)=\hU(x) J \ ,
\label{SigU}
\end{equation}
where
\begin{equation}
  \hU(x) = \exp\left(\frac{i\sqrt{2}\pi(x)}{F}\right)
  = \exp\left(\frac{i\sqrt{2}\pi_{\hat{a}}(x) X_{\hat{a}}}{F}\right) \ .
\label{Uhatrp}
\end{equation}
It can be verified that $\S(x)$
is symmetric (antisymmetric) for a real (pseudoreal) \irrep, as it should be.
As before, the vector gauge field is constructed from the unbroken generators,
while the axial one is constructed from the broken ones, \ie,
\begin{equation}
v_\m = v_{\m\tilde{a}}Q_{\tilde{a}}\ ,\qquad
a_\m = a_{\m\hat{a}}X_{\hat{a}}\ .
\label{vfields}
\end{equation}
  By using
the infinitesimal form of the transformation~(\ref{transrpSig}),
the covariant derivative is
\begin{eqnarray}
  D_\m \S
  &=& \partial_\m \hU J + i(v_\m+a_\m)\hU J + i \hU J (v_\m+a_\m)^T
\label{covderrp}\\
  &=& \left(\partial_\m \hU + i[v_\m,\hU] + i\{a_\m,\hU\} \right) J
\NON
  &\equiv& (D_\m \hU) J\ .
\nonumber
\end{eqnarray}

In writing down the chiral Lagrangian it will be convenient to use
notation which is as uniform as possible for all three cases.
To this end, we generalize Eq.~(\ref{SigU}) to the case of a complex \irrep\
by simply taking $J$ to be the $N\times N$ identity matrix in this case.
In all three cases: complex, real, and pseudoreal, the covariant derivative
is then given by Eq.~(\ref{covderrp}).

While we have discussed convenient choices for the matrix $J$
for the three types of \irreps, our results are valid more generally.
In particular, for the real and pseudoreal cases,
the derivation is valid for any matrix $J$ which satisfies the properties
discussed in the previous subsection.  For the convenience of the reader
we summarize them: $J$ must be an $N_w\times N_w$ real orthogonal matrix
with $\det\, J = 1$, and it should be symmetric (antisymmetric)
for the real (pseudoreal) case.

\subsection{\label{U1} Singlet axial symmetries}
In addition to the non-abelian flavor symmetry group,
we may apply to the fermions of each \irrep\ a flavor-singlet
axial transformation.  For Dirac fermions, this transformation is given by
\begin{equation}
  \j_i \to e^{-i\th\g_5}\j_i\ , \qquad \bj_i \to \bj_i e^{-i\th\g_5}\ .
\label{U1A}
\end{equation}
with a similar transformation for  Majorana fermions.
The corresponding $U(1)_A$ current is
\begin{equation}
  A_\m = \left\{ \begin{array}{ll}
    \sum_{i=1}^N \bj_i \g_\m \g_5 \j_i \ , \qquad & \mbox{Dirac fermions ,} \\
  \rule{0ex}{3ex}
    \sum_{I=1}^{N_w} \bJ_I \g_\m \g_5 \J_I \ , \qquad & \mbox{Majorana fermions .}
  \end{array} \right.
\label{U1AA}
\end{equation}
The Dirac version may be used for complex and pseudoreal \irreps,
whereas the Majorana version is used for real \irreps.\footnote{
  See App.~\ref{Majorana} for the definition of the Majorana fermion $\J$.
}

The individual $U(1)_A$ currents are anomalous
\begin{equation}
  \partial_\m A_\m = \frac{g^2}{32\p^2} N_w T\, F_{a\m\n} \tilde{F}_{a\m\n} \ ,
\label{FFdual}
\end{equation}
where the group-invariant $T$ is defined by
\begin{equation}
  \tr(T_a T_b)=T \d_{ab}\ ,
\label{tracenorm}
\end{equation}
where the $T_a$ are the generators of the gauge group
in the given \irrep, with $T=\half$ for the fundamental \irrep.
As usual, $N_w=2N$ in the case of Dirac fermions.

Consider an asymptotically free theory with fermions in $n$ different \irreps.
We will assume that if a given \irrep, $r$, is real,
the fermions are arranged as $N_{w,r}$ Majorana fields.
If $r$ is complex or pseudoreal, we assume that
the fermions may be assembled into $N_r=N_{w,r}/2$ Dirac fermions.\footnote{%
  If $r$ is a complex \irrep, we count both $r$ and its complex conjugate
  as the same \irrep, for the obvious reason that a Dirac fermion in a
  complex \irrep\ corresponds to two same-handedness Weyl fermions
  in the two complex conjugate \irreps.
}
In any such theory, only the overall
$U(1)_A$ transformation is anomalous, whereas $n-1$ linearly independent
combinations of the individual $U(1)_A$ currents are anomaly free.

\subsection{\label{twocoset} Parametrization of the coset fields: two \irreps}
From now on, we specialize to theories with fermions in two different \irreps.
The \irreps\ can be of the same type, \eg, both complex;
or they can be of different types, \eg, one complex \irrep\
and one real \irrep, as in the model of Ref.~\cite{Ferretti:2014qta}.
Out of the two flavor-singlet axial currents, we can make one linear
combination which is anomaly free.  Using indices $r,s,\ldots=1,2$
to label the two \irreps, the non-anomalous current is\footnote{%
  For Eq.~(\ref{U1nona}) to be true to all orders, $A_{r,\m}$ on the right-hand side
  should be the renormalized singlet axial current of the $r$-th \irrep.
}
\begin{equation}
  A_\m = \sum_r q_r A_{r,\m} \ ,
\label{U1nona}
\end{equation}
where, adopting a convenient normalization, the axial charges
of the two \irreps\ are
\begin{equation}
  q_1 = \frac{N_{w,2} T_2}{\sqrt{N_{w,1}^2 T_1^2 + N_{w,2}^2 T_2^2}} \ , \qquad
  q_2 = -\frac{N_{w,1} T_1}{\sqrt{N_{w,1}^2 T_1^2 + N_{w,2}^2 T_2^2}} \ .
\label{U1charge}
\end{equation}
The requirement that the current $A_\m$ be anomaly free only fixes
the ratio $q_1/q_2$.  As is usually the case for an abelian symmetry,
the overall normalization of the current is arbitrary.
Obviously, physics should not depend on this choice.
In App.~\ref{singlet} we discuss the choice of normalization
in a little more detail, showing that this is indeed the case.

For each \irrep,
the fermion condensate carries twice the axial charge of a single field.
It follows that the non-anomalous $U(1)_A$ is spontaneously broken, too.
To account for the corresponding NGB, we introduce a new effective field,
\begin{equation}
  \F(x) = \exp\left(\frac{i\z(x)}{\sqrt{2}F_\z}\right)\in U(1) \ ,
\label{Phi}
\end{equation}
with unit charge under $U(1)_A$.
The covariant derivative of this field is
\begin{equation}
  D_\m \F = \partial_\m \F + i\a_\m \F
  = i\F\left( \frac{\partial_\m \z}{\sqrt{2}F_\z} + \a_\m \right) \ ,
\label{covU1A}
\end{equation}
where $\a_\m$ is the (external) $U(1)_A$ gauge field.

In order to match all quantum numbers of the order parameters,
including their $U(1)_A$ charges, Eq.~(\ref{cmplxSig}) gets replaced by
\begin{eqnarray}
  \tr (\j_{L,i} \bj_{R,j}) & \leftrightarrow & \F^{2q} \S_{ij}  \ ,
\label{cmplxSigPhi}\\
  \tr (\j_{R,j} \bj_{L,i}) & \leftrightarrow & \F^{-2q} \S^*_{ij} \ ,
\nonumber
\end{eqnarray}
for the complex case, while~(\ref{rpSig}) gets replaced by
\begin{eqnarray}
  \tr (\x_{L,I} \x_{L,J}^T CS) & \leftrightarrow & \F^{2q} \S_{IJ}  \ ,
\label{rpSigPhi}\\
  \tr (\bx_{L,I}^T \bx_{L,J} CS^T ) & \leftrightarrow & \F^{-2q} \S^*_{IJ}  \ ,
\nonumber
\end{eqnarray}
for the real and pseudoreal cases.
In all cases, the chiral source $\c$ carries charge $+2q$.

\section{\label{ChL} Chiral Lagrangian}
We are now ready to write down the chiral Lagrangian for two
different \irreps, labeled by indices $r,s,\ldots=1,2$.
As before, when $r$ is a complex \irrep\ the flavor indices are
$i,j,\ldots=1,\ldots,N_r$, where $N_r$ is the number of Dirac fermions.
For real and pseudoreal \irreps, the flavor indices are
$I,J,\ldots=1,\ldots,N_{w,r}$, where $N_{w,r}$
is the number of Weyl fermions.\footnote{%
  Recall that $N_w$ is even for a pseudoreal \irrep.
}
To allow for more uniformity of our notation, we also introduce $n_r$,
which will be equal $N_r$ for a complex \irrep,
and to $N_{w,r}$ for real and pseudoreal \irreps.

\subsection{\label{LO} Leading order}
The leading-order (LO) Lagrangian consists of kinetic terms and mass terms,
\begin{equation}
  \cl_2 = \cl_k + \cl_m \ .
\label{L2}
\end{equation}
There is a separate kinetic term for each coset field,
\begin{equation}
  \cl_k = F_\z^2 (D_\m\F)^\dagger D_\m\F
  + \sum_r \frac{F_r^2}{4} \svev{(D_\m\S_r)^\dagger D_\m\S_r} \ ,
\label{LOk}
\end{equation}
where, from now on, we will use the notation $\svev{\cdots}$
to indicate tracing over the flavor indices.

The mass terms take the form
\begin{equation}
  \cl_m = -\sum_r \frac{F_r^2}{4} \svev{\c_r^\dagger U_r + U_r^\dagger \c_r} \ ,
\label{LOm}
\end{equation}
where we have introduced the product fields
\begin{equation}
  U_r(x) = \F(x)^{2q_r} \S_r(x) =  \F(x)^{2q_r} \hU_r(x) J_r \ .
\label{Ufield}
\end{equation}
The presence of $\F^{2q_r}$ is forced upon us because
$\c_r$ carries charge $2q_r$.
Using the results of Sec.~\ref{twocoset} it can be checked that $\cl_m$
is invariant under all the flavor symmetries,
including the non-anomalous $U(1)_A$.
With the external gauge fields turned on, the entire Lagrangian $\cl_2$
is thus invariant under local flavor transformations.

The LO Lagrangian is also invariant under an ``intrinsic'' parity symmetry
that acts simultaneously on all fields.
For a complex \irrep, the intrinsic parity is
\begin{eqnarray}
  && \S_r\to\S_r^\dagger \ , \qquad \c_r\to\c_r^\dagger \ ,
\label{parityc}\\
  && v_{r\m}  \to v_{r\m} \ , \qquad a_{r\m} \to -a_{r\m} \ ,
\nonumber
\end{eqnarray}
whereas for the other two cases it is
\begin{eqnarray}
  && \S_r\to s_r\S_r^\dagger \ , \qquad \c_r\to s_r\c_r^\dagger \ ,
\label{parityrp}\\
  && v_{r\m}  \to -v_{r\m}^T \ , \qquad a_{r\m} \to -a_{r\m}^T \ ,
\nonumber
\end{eqnarray}
where, as in Eq.~(\ref{rpSig}), $s_r=1$ ($s_r=-1$) for a real (pseudoreal) \irrep.
The transformation of the pion fields is $\p_r\to -\p_r$ for a complex \irrep,
and $\p_r\to -\p_r^T$ for real and pseudoreal \irreps.\footnote{%
  It follows from Eq.~(\ref{brkrp}) that if $X_a$ is a coset generator,
  so is $X_a^T$.
}
Finally, for the singlet sector, the intrinsic parity is
\begin{equation}
  \F\to\F^* \ , \qquad \a_\m \to -\a_\m \ .
\label{paritys}
\end{equation}

In order to develop the perturbative expansion we let the chiral sources
assume their ``expectation values,'' \ie, we set
\begin{equation}
  \c_r = 2m_r B_r J_r \ ,
\label{vevchi}
\end{equation}
where $m_r\ge 0$, and the allowed choices for $J_r$
are summarized in Sec.~\ref{coset}.
Using Eqs.~(\ref{covderrp}),~(\ref{Ufield}) and~(\ref{vevchi}) it can be checked
that the $J_r$ matrices completely drop out when the LO Lagrangian
is expressed in terms of the fields $\F$ and $\hU_r$.
We next use Eqs.~(\ref{Uhatc}),~(\ref{Uhatrp}) and~(\ref{Phi}) to
extract the quadratic part of the LO Lagrangian, obtaining
\begin{equation}
  \cl_2^{quad} = \half (\partial_\m\z \partial_\m\z + M_\z^2 \z^2)
  + \half \sum_r \svev{\partial_\m\p_r \partial_\m\p_r + M_r^2 \p_r^2} \ ,
\label{LOquad}
\end{equation}
where we have now turned off the external gauge fields.
The tree-level masses are
\begin{equation}
  M_r^2 = 2m_r B_r \ ,
\label{Msq}
\end{equation}
for the pions, and
\begin{equation}
  M_\z^2 = 2 \sum_r \frac{F_r^2}{F_\z^2}\, q_r^2 m_r B_r \vev{\id_r}
  = \sum_r \frac{F_r^2}{F_\z^2}\, q_r^2 M_r^2 n_r \ ,
\label{Meta}
\end{equation}
for the flavor singlet pseudoscalar $\z$,
where $n_r$ is defined at the beginning of Sec.~\ref{ChL}.
Note that $M_\z^2$ vanishes
only when the fermion masses of both \irreps\ vanish.
The tree-level ``quark flow'' propagators are obtained using
closure relations that we have collected in  App.~\ref{clsr}.
For a complex \irrep\ (dropping the \irrep's index $r$)
the propagator is\footnote{%
  In Eqs.~(\ref{propc}) and~(\ref{proprp}) the notation $\svev{\cdots}$ stands
  for an expectation value, not a flavor-index trace.
}
\begin{equation}
  \svev{\p_{ij}(x)\p_{k\ell}(y)}
  = \int\frac{d^4p}{(2\pi)^4}\, \frac{e^{ip(x-y)}}{p^2+M^2}
  \left(\delta_{i\ell}\delta_{jk}
  - \frac{1}{N}\,\delta_{ij}\delta_{k\ell}\right)\ .
\label{propc}
\end{equation}
For a real or pseudoreal \irrep, it is
\begin{equation}
  \svev{\p_{IJ}(x)\p_{KL}(y)}
  = \int\frac{d^4p}{(2\pi)^4}\, \frac{e^{ip(x-y)}}{p^2+M^2}
  \left(\frac{1}{2} \left(\delta_{IL}\delta_{JK} + J_{IK}J_{JL}\right)
  - \frac{1}{N_w}\,\delta_{IJ}\delta_{KL}\right)\ .
\label{proprp}
\end{equation}
This structure follows from the fact that the pion matrix obeys the relation
$\p=\half\left(\p+J\p^T J^T\right)$ and that it is traceless.

An advantage of the quark-flow Feynman rules is that the vertices
can be read off mechanically, and the coset structure is reflected only
in the above expressions for the propagators.  In particular,
this is the only place where one encounters the $J_r$ matrices once the
Lagrangian has been expressed in terms of the $\F$ and $\hU_r$ fields.

The expansion of the kinetic terms in the pion fields is standard.
In the mass terms, on the other hand, we encounter terms that depend
on both the pion and flavor singlet fields.  For example, the quartic part
of $\cl_m$ is
\begin{equation}
  \cl_m^{quart} = - \sum_r M_r^2 \left(
  \frac{n_r q_r^4 F_r^2}{12 F_\z^4}\, \z^4
  + \frac{q_r^2}{2 F_\z^2}\, \z^2 \vev{\p_r^2}
  + \frac{q_r}{3 F_\z F_r}\, \z \vev{\p_r^3}
  + \frac{1}{12 F_r^2}\, \vev{\p_r^4} \right) \ .
\label{Lm4}
\end{equation}

The $\z \vev{\pi_r^3}$ term appearing in this Lagrangian is a novel feature,
as, by itself, $\vev{\pi_r^3}$ violates intrinsic parity.
This interaction allows the decay $\z \rightarrow 3\pi$ to proceed
at tree level even when all fermion masses for a single \irrep\
are degenerate (if that mass is small enough),
unlike the similar decay $\eta \rightarrow 3\pi$ in QCD which requires
isospin violation to occur.

\subsection{\label{NLO} Next-to-leading order}
The next to leading order Lagrangian
\begin{equation}
  \cl_4 = \cl_s + \cl_d + \cl_\z \ ,
\label{L4}
\end{equation}
consists of several kinds of terms.
Following closely the classification of the QCD case
\cite{Gasser:1983yg,Gasser:1984gg},
we start with the single-trace terms
\begin{equation}
  \cl_s = \sum_r \left(L_{0r}P_{0r} - L_{3r} P_{3r}
                       + L_{5r} P_{5r} - L_{8r} P_{8r} -H_{2r} X_{2r}\right) \ ,
\label{Ls}
\end{equation}
where\footnote{%
  $P_{0r}$ is redundant for $N\le3$ for complex \irreps,
  or $N_w\le3$ for real and pseudoreal \irreps\ \cite{Golterman:2009kw}.
}
\begin{subequations}
\label{LsOP}
\begin{eqnarray}
  P_{0r} &=& \svev{(D_\mu \hU_r)^\dagger D_\nu \hU_r
                   (D_\mu \hU_r)^\dagger D_\nu \hU_r} \ ,
\label{LsOPa}\\
  P_{3r} &=& \svev{(D_\mu \hU_r)^\dagger D_\mu \hU_r
                   (D_\nu \hU_r)^\dagger D_\nu \hU_r} \ ,
\label{LsOPb}\\
  P_{5r} &=& \svev{(D_\mu \hU_r)^\dagger D_\mu \hU_r
                   (\chi_r^\dagger U_r + U_r^\dagger \chi_r)} \ ,
\label{LsOPc}\\
  P_{8r} &=& \svev{\chi_r^\dagger U_r \chi_r^\dagger U_r
                 + U_r^\dagger \chi_r U_r^\dagger \chi_r} \ ,
\label{LsOPd}\\
  X_{2r} & =& \svev{\chi_r^\dagger \chi}\ .
\label{LsOPe}
\end{eqnarray}
\end{subequations}
Here $H_{2r}$ is a ``high-energy'' constant, multiplying a contact term.
The minus signs in Eq.~(\ref{Ls}) and following, relative to
Refs.~\cite{Gasser:1983yg,Gasser:1984gg},
are present because we work in Euclidean metric while their metric is
Minkowski, and we want our results for observables to agree with theirs
in the single-representation case.
Note that, through $U_r$, some of these operators depend on the singlet
field $\F$.  Next, there are double-trace terms
\begin{equation}
  \cl_d = \sum_{rs} \left(-L_{1rs}P_{1rs} - L_{2rs}P_{2rs} + L_{4rs}P_{4rs}
  - L_{6rs}P_{6rs} - L_{7rs}P_{7rs} \right) \ ,
\label{Ld}
\end{equation}
where all low-energy constants (LECs) except $L_{4rs}$ are symmetric under
$r\leftrightarrow s$, and
\begin{subequations}
\label{LdOP}
\begin{eqnarray}
  P_{1rs} &=& \svev{(D_\mu \hU_r)^\dagger D_\mu \hU_r}
              \svev{(D_\nu \hU_s)^\dagger D_\nu \hU_s} \ ,
\label{LdOPa}\\
  P_{2rs} &=& \svev{(D_\mu \hU_r)^\dagger D_\nu \hU_r}
              \svev{(D_\mu \hU_s)^\dagger D_\nu \hU_s} \ ,
\label{LdOPb}\\
  P_{4rs} &=& \svev{(D_\mu \hU_r)^\dagger D_\mu \hU_r}
              \svev{\chi_s^\dagger U_s + U_s^\dagger \chi_s} \ ,
\label{LdOPc}\\
  P_{6rs} &=& \svev{\chi_r^\dagger U_r + U_r^\dagger \chi_r}
              \svev{\chi_s^\dagger U_s + U_s^\dagger \chi_s} \ ,
\label{LdOPd}\\
  P_{7rs} &=& \svev{\chi_r^\dagger U_r - U_r^\dagger \chi_r}
              \svev{\chi_s^\dagger U_s - U_s^\dagger \chi_s} \ .
\label{LdOPe}
\end{eqnarray}
\end{subequations}
Finally, there are additional terms that involve the singlet field's
two-derivative operator
\begin{equation}
  \cl_\z = L'_0P'_0
  - \sum_r \left(L'_{1r}P'_{1r} + L'_{2r} P'_{2r} + L'_{3r} P'_{3r} \right) \ ,
\label{Leta}
\end{equation}
where
\begin{subequations}
\label{LetaOP}
\begin{eqnarray}
  P'_0 &=& \left((D_\mu\F)^\dagger D_\mu\F \right)^2 \ ,
\label{LetaOPa}\\
  P'_{1r} &=& \svev{(D_\mu \hU_r)^\dagger D_\mu \hU_r}
                    (D_\mu\F)^\dagger D_\mu\F \ ,
\label{LetaOPb}\\
  P'_{2r} &=& \svev{(D_\mu \hU_r)^\dagger D_\nu \hU_r}
                    (D_\mu\F)^\dagger D_\nu\F \ ,
\label{LetaOPc}\\
  P'_{3r} &=& \svev{\chi_r^\dagger U_r + U_r^\dagger \chi_r}
                    (D_\mu\F)^\dagger D_\mu\F \ .
\label{LetaOPd}
\end{eqnarray}
\end{subequations}

\section{\label{1loop} Next-to-leading order results}
In this section we will present the NLO corrections
for the masses and the decay constants of the (pseudo) NGBs,
and for the condensates.
Since the calculations leading to these results are straightforward,
we will not give any details.  We have cross-checked all applicable
single-representation results in these formulas (\ie,
analytic terms and chiral logarithms which do not involve $\z$)
against the corresponding NLO results in the literature \cite{Bijnens:2009qm}.

\subsection{\label{masses} Pseudoscalar masses}
The inverse propagator takes the general form
\begin{equation}
  p^2 + M^2 + \G(p^2) \ ,
\label{invprop}
\end{equation}
where $M^2$ is the tree-level mass, and $\G(p^2)$ is the NLO
self-energy.   The physical mass-squared $M^2_{phys}
=M^2+\d M^2$ is equal to the value of $-p^2$ for which this vanishes.
At NLO, we may set $p^2=-M^2_{phys}\to-M^2$ in $\G(p^2)$, and we obtain
\begin{equation}
  \d M^2 = \G(-M^2) \ .
\label{dMsq}
\end{equation}
We first consider the pions.  The NLO correction can be expressed as
\begin{equation}
  \d M^2_r = \d M^2_{r,an} +  \d M^2_{r,\p} +  \d M^2_{r,\z} \ .
\label{dMpi}
\end{equation}
The origin of the various terms is the following.
$\d M^2_{r,an}$ is the analytic contribution from the NLO Lagrangian,
given by
\begin{equation}
  \d M^2_{r,an} = \frac{8M_r^2}{F_r^2} \left(
  (2L_{8r}-L_{5r}) M_r^2
  + \sum_s (2L_{6rs}-L_{4rs}) M_s^2 n_s \right) \ .
\label{dMpian}
\end{equation}
$\d M^2_{r,\p}$ is the usual non-analytic contribution from a pion tadpole,
which arises from a single quartic vertex of the LO Lagrangian.
It is given by \cite{Bijnens:2009qm}
\begin{equation}
 \d M^2_{r,\p} = M^2_r C_r \D_r\ ,
\label{dMpipi}
\end{equation}
where $C_r= 1/n_r$ for complex, $-1/2+1/n_r$ for real,
and $1/2 + 1/n_r$ for pseudoreal representations, and with
\begin{equation}
  \D_r = \frac{M_r^2}{16\p^2 F_r^2}\,\log\frac{M_r^2}{\m^2} \ .
\label{Delta}
\end{equation}
We are using the standard ChPT subtraction scheme
in which the tadpole is given entirely by the logarithm and its constant terms
are absorbed into the renormalized $L_i$'s \cite{Gasser:1984gg}.

Finally $\d M^2_{r,\z}$ is a similar non-analytic contribution involving
a $\z$ tadpole, which arises from the second term on the right-hand side
of Eq.~(\ref{Lm4}).  Explicitly
\begin{eqnarray}
  \d M^2_{r,\z} &=& -q_r^2 M_r^2 \D_\z \ ,
\label{dMpieta}\\
  \D_\z &=& \frac{M_\z^2}{16\p^2 F_\z^2}\,\log\frac{M_\z^2}{\m^2} \ .
\label{Deltaeta}
\end{eqnarray}

For the one-loop correction to the mass of the pseudoscalar singlet
we similarly find
\begin{equation}
  \d M^2_\z = \d M^2_{\z,an} +  \d M^2_{\z,\z} +  \d M^2_{\z,\p} \ .
\label{dMeta}
\end{equation}
The analytic contribution is
\begin{eqnarray}
  \d M^2_{\z,an} &=& \frac{1}{F_\z^2} \sum_r M_r^2 n_r
  \left( 16L_{8r} M_r^2 q_r^2 + 2 L'_{3r} M_\z^2 \right)
\label{dMetaan}\\
  && +\frac{1}{F_\z^2} \sum_{rs} M_r^2 M_s^2 n_r n_s
  \left( 8L_{6rs} (q_r^2+q_s^2) + 16L_{7rs} q_r q_s \right)\ .
\nonumber
\end{eqnarray}
The non-analytic contribution from a $\z$ tadpole is
\begin{equation}
  \d M^2_{\z,\z} = - \sum_r M_r^2 n_r q_r^4 \frac{F_r^2}{F_\z^2}  \Delta_\z\ ,
\label{dMetaeta}
\end{equation}
and the non-analytic contribution from pion tadpoles is
\begin{equation}
  \d M^2_{\z,\p} = -\sum_r D_r M_r^2 q_r^2  \frac{F_r^2}{F_\z^2} \D_r\ ,
\label{dMetapi}
\end{equation}
where the dimensionality of the coset, $D_r$,
is equal to $n_r^2-1$ for a complex representation,
$\half n_r(n_r+1)-1$ for a real representation,
and $\half n_r(n_r-1)-1$ for a pseudoreal representation.

\subsection{\label{decay} Decay constants}
As in the case of the pion mass, we write
\begin{equation}
\d F_r = \d F_{r,an} + \d F_{r,\p} + \d F_{r,\z}\ .
\label{dfpi}
\end{equation}
We find that
\begin{eqnarray}
\label{dfpilist}
\d F_{r,an}&=&4F_r\left(L_{5r}\frac{M_r^2}{F_r^2}
+\sum_s L_{4rs}n_s\frac{M_s^2}{F_r^2}\right)\ ,\\
\d F_{r,\p}&=&- \half F_r n_r\D_r\ ,\nonumber\\
\d F_{r,\z}&=&0\ .\nonumber
\end{eqnarray}
There are no loop contributions to $F_\z$, and we find a purely analytic result
\begin{equation}
\label{dfeta}
\d F_\z=-F_\z\sum_r L'_{3r}n_r\frac{M_r^2}{F_\z^2}\ .
\end{equation}

\subsection{\label{cond} Condensates}
The condensate $\S_r\equiv\svev{\bj_r \j_r}$ per flavor of \irrep\ $r$
is defined by
\begin{equation}
\label{defcond}
\S_r=-\frac{1}{n_r}\frac{\partial \log Z}{\partial m_r}\ .
\end{equation}
To leading order this yields $\S_r^0=-F_r^2 B_r$, using Eq.~(\ref{LOm}).  At NLO,
we again define
\begin{equation}
\label{eq:cond}
\d\S_r=\d\S_{r,an}+\d\S_{r,\p}+\d\S_{r,\z}\ ,
\end{equation}
for the analytic, pion-loop and $\z$-loop contributions.   A straightforward
calculation finds
\begin{eqnarray}
\label{condlist}
\d\S_{r,an}&=&4\S_r^0\left((2L_{8r}+H_{2r})\frac{M_r^2}{F_r^2}
+4\sum_s L_{6rs}n_s\frac{M_s^2}{F_r^2}\right)\ ,\\
\d\S_{r,\p}&=&-\S_r^0\, \frac{D_r}{n_r}\,\D_r\ ,\nonumber\\
\d\S_{r,\z}&=&-\S_r^0 q_r^2\D_\z\ ,\nonumber
\end{eqnarray}
where $D_r$ was defined below Eq.~(\ref{dMetapi}).

\section{\label{conc} Conclusion}
In this paper, we developed Chiral Perturbation Theory for a vector-like
gauge theory with fermions transforming in two different irreducible
representations of the gauge group.  We considered fermions in any type
of representation of the gauge group, complex, real or pseudoreal.  We
assumed that bilinear condensates develop for each of these fermions,
breaking the flavor symmetry of each fermion species spontaneously.

The low-energy effective field theory contains Nambu--Goldstone bosons
associated with the vacuum manifolds for each of the two condensates.
In addition, it contains one more singlet Nambu--Goldstone boson, because
a linear combination of the two axial $U(1)$ symmetries remains non-anomalous.
The two fermion condensates both break this singlet axial $U(1)$, and the
associated axial current thus creates a singlet Nambu--Goldstone boson from the
vacuum.

We allowed for degenerate masses for each fermion species;
of course, they are not degenerate between different irreducible
representations.  This turns the Nambu--Goldstone bosons into massive
pseudo Nambu--Goldstone bosons.
The (tree level) mass-squared of the singlet Nambu--Goldstone boson
is a linear combination of the masses of the two fermion species,
and it is thus not possible to give this Nambu--Goldstone
boson a mass without giving at least one of the non-singlet
Nambu--Goldstone boson multiplets a mass.  We presented next-to-leading
order results for all meson masses, decay constants, and the two
condensates.  It should be straightforward to generalize the
framework of this paper to a theory with more than two different types
of fermions.

We can imagine two potential uses for these results. The first, as
mentioned in the Introduction, is that theories as considered here have
applications in models for physics beyond the standard model. We have
not seen a systematic construction of the chiral Lagrangian for theories
with more than one representation of fermions presented to date, so our
results might be a resource for model builders. From this perspective,
we think that the most interesting aspect of these systems is the
appearance of the additional $U(1)$ Nambu--Goldstone boson.
The non-singlet Nambu--Goldstone bosons come in degenerate-mass multiplets,
if degenerate masses are given to the fermions in the underlying theory.
On the other hand, the singlet appears as a somewhat isolated state,
particularly if the masses of the two non-singlet multiplets are
somewhat separated.  This is a distinctive feature in the context
of \eg\ composite Higgs models, where new resonances tend to appear
with large multiplicity and similar masses.
Unusually for a Nambu-Goldstone boson, the singlet
can decay at tree level as $\z \rightarrow 3\pi$ even when all fermion
masses for each representation are degenerate, which may have
interesting phenomenological consequences in some theories.

The second potentially useful application of these results is as a
theoretical benchmark for the interpretation of lattice simulations
relevant for various extensions of the Standard Model \cite{DeGrand:2015zxa}.
Three of us are
involved in such an effort \cite{toappear}.  The interesting physics
issues are whether such a system is confining and chirally broken, and
if so, how the dimensionful parameters (decay constants, masses,
condensates) for the different representations are related to each
other. Is it possible that there are ranges of bare parameters in which
the fermions in one representation condense, while those in the others
do not?  Speculations about such behavior are long-standing
(see \eg\ Ref.~\cite{Raby:1979my}).  Of course, the results of this paper
apply only in the case that the fermions in both representations condense.

Seeing the additional $U(1)$ Nambu--Goldstone boson in a lattice
calculation might be difficult. One would have to measure
``quark-disconnected'' diagrams like those used in the measurement of
the $\eta'$ mass in QCD. An elaborate multi-channel analysis along the
lines of Ref.~\cite{Dudek:2011tt} might be needed to observe them. The
ordinary pions will be easier to study. There, the interesting physics
is the dependence of the squared mass $M_r^2$ of a pseudoscalar, or of
its decay constant $F_r$, on the mass of a fermion in a representation
$s \ne r$ when the mass of a fermion in representation $r$ is fixed.

\vspace{3ex}
\noindent {\bf Acknowledgments}
\vspace{3ex}

MG and YS would like to thank the Department of Physics of the University of
Colorado, and YS would like to thank the Department of Physics  and
Astronomy of San Francisco State University, for hospitality.
The research of TD, MG, and EN is supported in part by the
US Department of Energy under grants DE-SC0010005 (TD and EN)
and DE-FG03-92ER40711 (MG). YS is supported by the Israel Science Foundation
under grant no.~449/13.  Brookhaven National Laboratory is supported
by the U.~S.~Department of Energy under contract DE-SC0012704.

\appendix
\section{\label{Majorana} Majorana fermions}
When the Dirac fermions belong to a real \irrep\ we may alternatively
introduce Majorana fermions $\J_I$, $I=1,\ldots,2N$, where
\begin{eqnarray}
  \J_{L,i} &=& \j_{L,i} \ ,
\label{Psi}\\
  \J_{L,N+i} &=& SC\,\bj^T_{R,i} \ ,
\NON
  \J_{R,i} &=& SC\,\bj^T_{L,i} \ ,
\NON
  \J_{R,N+i} &=& \j_{R,i} \ ,
\nonumber
\end{eqnarray}
and where $i=1,\ldots,N$ as before.
Defining
\begin{equation}
  \bJ \equiv \J^T CS \ ,
\label{bJ}
\end{equation}
the Lagrangian becomes
\begin{equation}
\cl=\half\sum_{I=1}^{2N}\bJ_I\Dslash\J_I \ .
\label{lagreal}
\end{equation}
For a real \irrep,
the number of Majorana (or Weyl) fermions $N_w$ is also allowed to be odd,
in which case we can use a Majorana (or Weyl) basis, but not a Dirac basis.
Correspondingly,
allowing the range of summation in Eq.~(\ref{lagreal}) to be an arbitrary
positive integer $N_w$, the flavor symmetry acts as
\begin{equation}
\J\to\left(P_L g+P_R g^*\right)\J\ ,
\label{SUNm}
\end{equation}
where $g\in SU(N_w)$.  For $N_w=2N$, it can be checked that Eq.~(\ref{SUNm})
agrees with Eq.~(\ref{SU2Nw}).

In terms of the Majorana fields, we have (compare Eq.~(\ref{bilchi}),
and recall $S^T=S$)
\begin{equation}
  \J_I^T CS \J_J = \J_J^T CS \J_I \ .
\label{bilmaj}
\end{equation}
Moreover, since $C\g_5=\g_5^TC$, this remains true if the same
chiral projector is inserted on both sides of the equation.
All these bilinears are therefore symmetric on their flavor indices,
as expected.

\section{\label{clsr} Projectors}
In writing down the expressions for the tree-level propagators
we use that, with the normalization~(\ref{tnorm}),
the projector on the traceless hermitian generators of $SU(N)$ is
\begin{equation}
  P_{IJKL} \equiv T_{aIJ} T_{aKL}
  = \d_{IL} \d_{JK} - \frac{1}{N} \d_{IJ}\d_{KL} \ .
\label{projT}
\end{equation}
Splitting it into a projector on the space spanned by the $Q$'s
of Eq.~(\ref{unbrkrp}) and the $X$'s of Eq.~(\ref{brkrp}), we have
$P = P^Q + P^X$ where
\begin{eqnarray}
  P^Q_{IJKL} &\equiv& Q_{\tilde{a}IJ}Q_{\tilde{a}KL}
  = \half \Big( \d_{IL} \d_{JK} - J_{IK} J_{JL} \Big) \ ,
\label{PQ}\\
  P^X_{IJKL} &\equiv& X_{\hat{a}IJ}X_{\hat{a}KL}
  =
  \half \Big( \d_{IL} \d_{JK} + J_{IK} J_{JL} \Big)
  - \frac{1}{N} \d_{IJ}\d_{KL} \ .
\label{PX}
\end{eqnarray}
These results can be proved by rewriting Eqs.~(\ref{unbrkrp}) and Eq.~(\ref{brkrp}) as
\begin{eqnarray}
   Q_{\tilde{a}} &=& \half \Big( Q_{\tilde{a}} - J Q_{\tilde{a}}^T J^T \Big) \ ,
\label{unbrkrp3}\\
   X_{\hat{a}} &=& \half \Big(  X_{\hat{a}} + J X_{\hat{a}}^T J^T \Big) \ ,
\label{brkrp3}
\end{eqnarray}
which are valid for any real orthogonal matrix $J$.

\section{\label{singlet} Normalization of the singlet axial current}
In Eq.~(\ref{U1charge}) we chose a particular normalization of the charges
$q_1$ and $q_2$, and thus a particular normalization of the non-anomalous
singlet axial current defined in Eq.~(\ref{U1nona}).
Furthermore, the singlet's decay constant $F_\z$ is defined by\footnote{%
  $F_\z$ is the decay constant in the full chiral limit,
  where the masses of all fermions in the underlying theory vanish.
}
\begin{equation}
\langle 0|A_\m(x)|\z\rangle = ip_\m\,\sqrt{2}F_\z\,e^{ipx}\ ,
\label{fetadef}
\end{equation}
in analogy with Eq.~(\ref{fpidef}).   $F_\z$ will show up in, for example,
the $\z$ decay rate, and it is therefore instructive to check that
$\z$ physics  is not affected by the choice of normalization.

If we change the normalization of $A_\m$ by a factor $\l$,
it follows from Eq.~(\ref{fetadef}) that the decay constant
is rescaled as $F_\z\to\l F_\z$, and from Eq.~(\ref{U1nona}) that $q_r\to\l q_r$.
If we now turn off the external gauge fields, and reexpress the LO lagrangian
in terms of the $\z$ field, the result will depend only on the
ratios $q_r/F_\z$, which are invariant.   This is true, in particular,
for the factor of $\F^{2q_r}$ that occurs in Eq.~(\ref{LOm}).
Other concrete examples are provided by the LO singlet mass~(\ref{Meta}),
and the interaction vertices~(\ref{Lm4}).

Proceeding to the NLO results we have calculated, corrections to masses,
to the pion decay constants, and to the condensates,
should be invariant under the rescaling, whereas corrections to the singlet
decay constant should scale in the same way as $F_\z$ itself.\footnote{%
  Note that in order to probe the singlet decay constant we need
  to turn back on the singlet axial gauge field.
}
One can then read off from our NLO results how the NLO LECs should scale.
The unprimed NLO LECs are invariant, while the primed ones rescale
as $L'_{ir}\to \l^2 L'_{ir}$, $i=1,2,3$, and $L'_0\to \l^4 L'_0$.
Alternatively, these scaling rules can be inferred from the contribution
of these NLO terms to the singlet axial current,
in comparison with the LO term following from Eq.~(\ref{LOk}).   Indeed,
in our explicit NLO results, $L'_{3r}$ always appears in the combination
$L'_{3r}/F_\z^2$, which is independent of $\l$.

We conclude with one more example.  In the context of composite Higgs models,
when the couplings to Standard-Model gauge fields are turned on,
the $A_\m$ current becomes anomalous, and $\z$ develops
anomaly-induced couplings to pairs of Standard-Model vector bosons.
If, for example, we turn on electromagnetism, this anomaly takes the form
\begin{equation}
\partial_\m A_\m = e^2 F_{\m\n}\tilde{F}_{\m\n} \sum_r q_r c_r \ ,
\label{EMan}
\end{equation}
where $F_{\m\n}$ is the electromagnetic field strength, and
where $c_r$ is a weighted sum over the squared electric charges
of the fermions that belong to the $r$-th \irrep.  Using that
\begin{equation}
A_\m=\sqrt{2}F_\z\partial_\m \z+ \mbox{higher\ orders}\ ,
\label{eftaxial}
\end{equation}
we find what is essentially the $\z$ equation of motion to this order, \ie,
\begin{equation}
\bo\z=\frac{e^2}{\sqrt{2}}\,F_{\m\n}\tilde{F}_{\m\n}
\sum_r \frac{q_r}{F_\z}\, c_r \ .
\label{EOM}
\end{equation}
Again, only the ratio $q_r/F_\z$ appears, implying that the decay rate
is independent of the arbitrary choice of normalization of the
singlet axial current $A_\m$.

\vspace{5ex}

\end{document}